\documentclass[final,5p,times,twocolumn]{elsarticle}

\usepackage{comment}
\graphicspath{ {images/} }
\usepackage{amsfonts}
\usepackage{mathrsfs}
\usepackage{amssymb}
\usepackage{amsthm}
\usepackage{graphicx}
\usepackage{xspace}
\usepackage{subcaption}
\usepackage{footnote}
\usepackage{mathtools}
\usepackage{url}
\usepackage{chngcntr}
\usepackage{listings, color, caption}
\usepackage{threeparttable}
\usepackage{setspace}
\usepackage{multirow}
\usepackage{multicol}
\usepackage{pifont}
\usepackage{soul}
\usepackage{boxedminipage}
\usepackage{booktabs}
\usepackage{multirow}
\usepackage{hyperref}
\usepackage[ruled]{algorithm2e}
\usepackage[noend]{algpseudocode}
\usepackage{indentfirst}

\journal{High-Confidence Computing}

\begin{document}

\begin{frontmatter}

\title{JFinder: A Novel Architecture for Java Vulnerability Identification Based Quad Self-Attention and Pre-training Mechanism}

\author[label1]{Jin Wang\fnref{first}}
\author[label1]{Zishan Huang\fnref{first}}
\author[label1]{Hui Xiao}
\author[label1]{Yinhao Xiao\corref{cor}}

\ead{20191081@gdufe.edu.cn}
\cortext[cor]{Corresponding author}

\fntext[first]{These authors contributed equally to this work}

\affiliation[label1]{organization={School of Information Science},
	addressline={Guangdong University of Finance and Economics}, 
	city={Guangzhou},
	postcode={510320}, 
	state={Guangdong},
	country={China}}

\begin{abstract}
	Software vulnerabilities pose significant risks to computer systems, impacting our daily lives, productivity, and even our health. Identifying and addressing security vulnerabilities in a timely manner is crucial to prevent hacking and data breaches. Unfortunately, current vulnerability identification methods, including classical and deep learning-based approaches, exhibit critical drawbacks that prevent them from meeting the demands of the contemporary software industry. To tackle these issues, we present JFinder, a novel architecture for Java vulnerability identification that leverages quad self-attention and pre-training mechanisms to combine structural information and semantic representations. Experimental results demonstrate that JFinder outperforms all baseline methods, achieving an accuracy of 0.97 on the CWE dataset and an F1 score of 0.84 on the PROMISE dataset. Furthermore, a case study reveals that JFinder can accurately identify four cases of vulnerabilities after patching.
\end{abstract}
\begin{keyword}
	Software Vulnerabilities \sep Security Risks \sep Vulnerability Identification Methods \sep Pre-training \sep Structural Informatin \sep Self-Attention



\end{keyword}

\end{frontmatter}

\section{Introduction}
\label{sec:introduction}
As modern software continues to increase in functionality, the likelihood of vulnerabilities also grows. These vulnerabilities pose significant risks to cybersecurity, with implications for individuals, the healthcare industry, and industrial production. For individuals, such vulnerabilities can result in the leak of sensitive information, leading to identity theft and fraud. In the healthcare sector, cybersecurity breaches may entail the theft of health information, ransomware attacks on hospitals, and even attacks on implanted medical devices. In industrial production, software vulnerabilities can introduce corresponding vulnerabilities in products reliant on that software, as exemplified by the Log4j2 vulnerability. Apache Log4j2 is susceptible to remote code execution (RCE) attacks\footnote{https://logging.apache.org/log4j/2.x/security.html}, wherein an attacker with the ability to modify logging configuration files can create malicious configurations using the JDBC Appender and a data source referencing a JNDI URI, enabling remote code execution. The Log4j2 vulnerability impacted tens of thousands of products but was swiftly addressed, preventing a catastrophic cyber event.\par
The identification of vulnerabilities is crucial for ensuring system security and timely remediation of security flaws, thus protecting against hacking and data breaches. However, vulnerability detection can be a laborious and challenging process. Researchers have been continuously exploring methods to automate this task. Initially, researchers manually identified features and employed machine learning to ascertain the existence of vulnerabilities, but this approach proved time-consuming. Subsequently, deep learning techniques were utilized to automatically detect vulnerability features and classify them. Many of these methods involve the use of graph neural networks to identify vulnerability patterns, leading to the development of new graph neural network types (e.g., Graph Convolution Networks, Graph Attention Networks, and Graph Autoencoders). Some approaches rely on structural information (e.g., Abstract Syntax Tree, Control Flow Graph, and Data Flow Graph) to construct graphs, while others solely employ semantic information of codes for embedding. However, these methods have not demonstrated satisfactory performance on real software vulnerability datasets and remain unsuitable for industrial application.\par
To address these issues, we propose JFinder, a novel architecture for Java vulnerability identification leveraging structural information with MetaPaths, a quad self-attention layer, and a pre-trained programming language model. We utilize a third-party library to obtain the Abstract Syntax Tree (AST), Control Flow Graph (CFG), and Data Flow Graph (DFG) of source code as structural information. We derive the Code Snippet Sequence (CSS) using a pre-trained model, UniXcoder~\cite{guo2022unixcoder}, a transformer-based language model designed for natural language processing tasks in the software development domain, trained on an extensive corpus of source code and natural language text related to software development. UniXcoder enables the conversion of program language into a feature matrix, providing accurate semantic representations of code snippets as it is trained on program language. JFinder then feeds semantic and structural information into convolutional networks and multilayer perceptrons to predict vulnerability presence. Overall, JFinder uniquely combines semantic and structural information to comprehensively analyze the execution process from multiple perspectives. We have implemented JFinder as an open-source project on GitHub\footnote{https://github.com/WJ-8/JFinder}. We evaluated JFinder on CWE and PROMISE datasets, comprising a total of 20,402 code snippets, of which 7,355 were vulnerable. Experimental results indicate that JFinder achieved outstanding performance on the CWE dataset, with an accuracy rate of 97\%. On the PROMISE dataset, JFinder attained an industrially viable level with F1 scores reaching 0.83. We also conducted case studies with four cases; after vulnerabilities were addressed, JFinder no longer identified these cases as vulnerable. The case study results demonstrated the capacity of JFinder to uncover robust vulnerability patterns and provide insights. Our contributions are threefold: \par
\begin{itemize}
	\item We propose a novel architecture for java vulnerability identification, JFinder, which combines structural information and semantic information from a code snippet. We open the source of JFinder in Github.
	\item We have conducted a large number of experiments and compared them with recent excellent methods. The results show that JFinder outperforms all baseline methods and the results are satisfactory.
	\item We conduct case studies to explore the robustness and intelligence level of JFinder. Experience results show that JFinder understands the meaning of the vulnerabilities in depth.
\end{itemize}
\textbf{Paper Organization.} 
The rest of the paper is organized as follows. Section~\ref{sec:background} presents the recently advanced background knowledge of our approach. Section~\ref{sec:approach} details the design and technical components of the JFinder framework. Section~\ref{sec:implementation} demonstrates our implementation of JFinder. Section~\ref{sec:eval} reports our evaluation results and case studies on JFinder. Section~\ref{sec:related} outlines the most related work.  Section~\ref{sec:conclusion} concludes the paper with a future research discussion.
\section{Background}
\label{sec:background}
In this section, we present the background knowledge of some recently advanced technologies utilized by JFinder.

\subsection{Pre-trained Models}
Pre-trained models are machine learning models that have already undergone training for certain tasks. These models can significantly reduce training time and achieve better results without requiring large amounts of data. Notable pre-trained models such as BERT and GPT exhibit exceptional performance in natural language processing (NLP) and serve as milestones in the field of artificial intelligence. Due to their complex pre-training objectives and large parameter count, pre-trained models can effectively capture knowledge from vast amounts of labeled and unlabeled data. By storing knowledge in numerous parameters and fine-tuning specific tasks, the rich knowledge encoded in these parameters can benefit various downstream tasks, as demonstrated by experimental validation and empirical analysis. Pre-training mechanisms enable models to learn generic linguistic expressions by leveraging substantial volumes of unlabeled text. Pre-trained models can be adapted to downstream tasks by adding one or two specific layers, providing a good initialization and preventing the need to train downstream models from scratch. This approach improves performance on small datasets, reducing the requirement for a large number of labeled instances. As deep learning models with many parameters tend to overfit on small datasets, pre-training serves as a form of regularization by providing a good initialization and avoiding overfitting.

\subsection{Self-Attention Mechanism}
The self-attention mechanism constitutes a crucial component of pre-trained models commonly employed for NLP tasks, such as the Transformer. This mechanism determines the relationships between words by focusing on the input data's positions, resulting in more efficient text representations. The self-attention mechanism encodes the contextual information of an entire text in each word's semantic representation. Its advantages include capturing long-distance dependencies in text and effectively handling variable-length sequences. Owing to its outstanding performance in NLP, the self-attention mechanism has become a popular research topic in the field.

In our implementation, we utilize a multi-head self-attention mechanism. This approach replicates a single attention head into multiple ones, applying them to different data positions separately to obtain more semantic information. Each head independently learns distinct contextual information, enhancing the model's generalization capabilities. This method is widely used in Transformer models and has achieved satisfactory practical results.

\subsection{Word Embedding}
Traditional machine learning methods often struggle to process textual data directly, necessitating appropriate techniques to convert text data into numerical data, which introduces the concept of word embedding. Word embedding encompasses language modeling and representation learning techniques in NLP, serving as an early pre-training technique. It refers to embedding a high-dimensional space, with dimensions equal to the total number of words, into a continuous vector space of much lower dimensionality. In this space, each word or phrase is mapped to a vector of real numbers, representing a specific concept through a distributed representation.

\subsubsection{Bag-of-words}
In information retrieval, the bag-of-words model assumes that a text is a collection of words or word combinations, disregarding word order, grammar, and syntax. The model considers the occurrence of each word in a text to be independent of the occurrences of other words. Common applications include one-hot encoding and N-gram techniques. Although the model is easy to understand and implement, it has several drawbacks. Careful vocabulary design is crucial, particularly to manage size and avoid sparse context representations. Disregarding word order neglects the context and meaning of words in a document.

\subsubsection{Context-independent with machine learning}
Context-independent machine learning approaches do not consider contextual information in the learning process. They assume input data is independent of other information, focusing solely on the current input during model training. These approaches are commonly used for text classification, with popular models including word2vec, fastText, and glove.
\subsubsection{Context-dependent and transformer-based}
Context-dependent and transformer-based methods represent context-sensitive approaches in which the same word is represented differently depending on the context. These methods obtain contextual information by using a transformer model to compute the relationships of the input data. Transformer models incorporate attention mechanisms that learn long-range dependencies, making them better suited for natural language processing (NLP) tasks.
\section{Approach}
\label{sec:approach}
\begin{figure*}[htbp]
	\includegraphics[width=190mm]{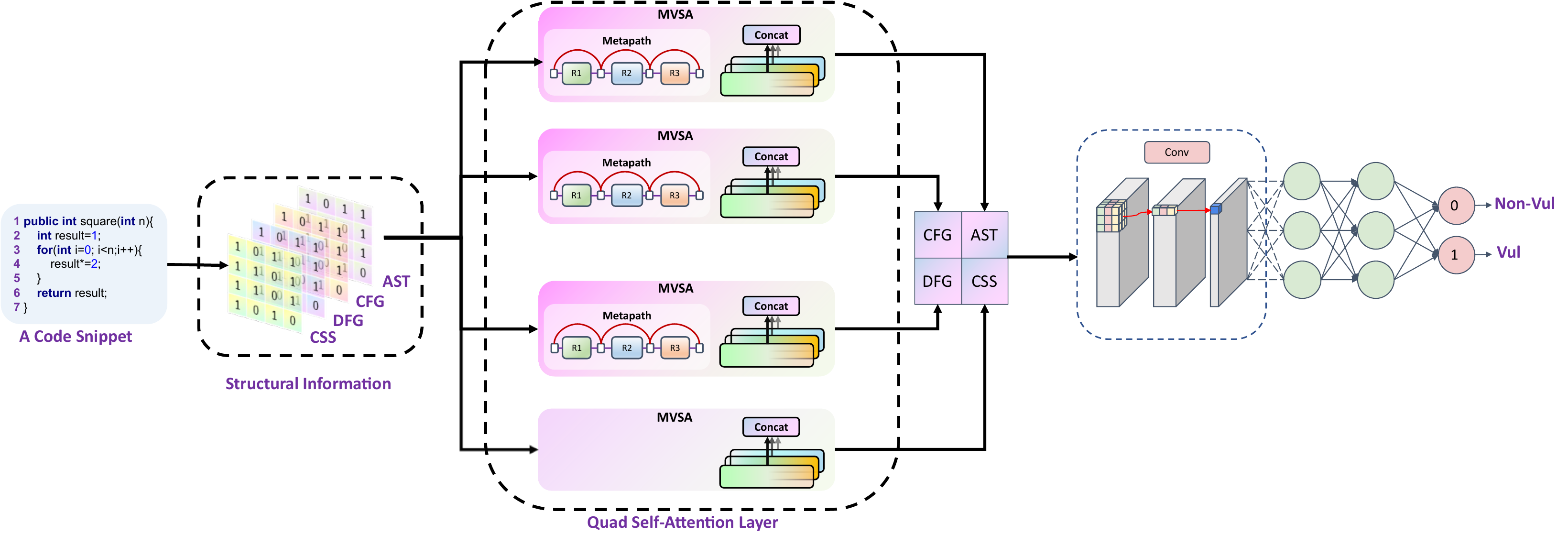}
	\centering
	\caption{Illustration of the JFinder model}
	\label{fig:model}
\end{figure*}

In this section, we detail the design of JFinder, illustrated in Fig.~\ref{fig:model}. The model workflow is as follows: we input a code snippet and parse it to obtain the AST, CFG, DFG, and CSS matrices. These matrices are then fed into the Quad Self-Attentive Layer and merged into a single matrix. Finally, a convolutional neural network and a multilayer perceptron predict whether the input code snippet is vulnerable or not.
\subsection{Structural Information}

\begin{figure*}[htp]
	\centering
	\begin{minipage}[t]{0.22\linewidth}
		\centering
		\includegraphics[width=\textwidth]{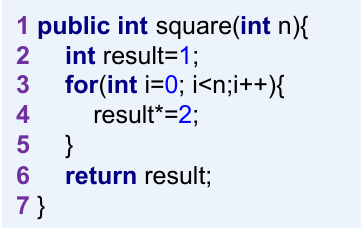}
		\caption{A Code Example}
		\label{fig:code}
	\end{minipage}
	\begin{minipage}[t]{0.7\linewidth}
		\centering
		\includegraphics[width=\textwidth]{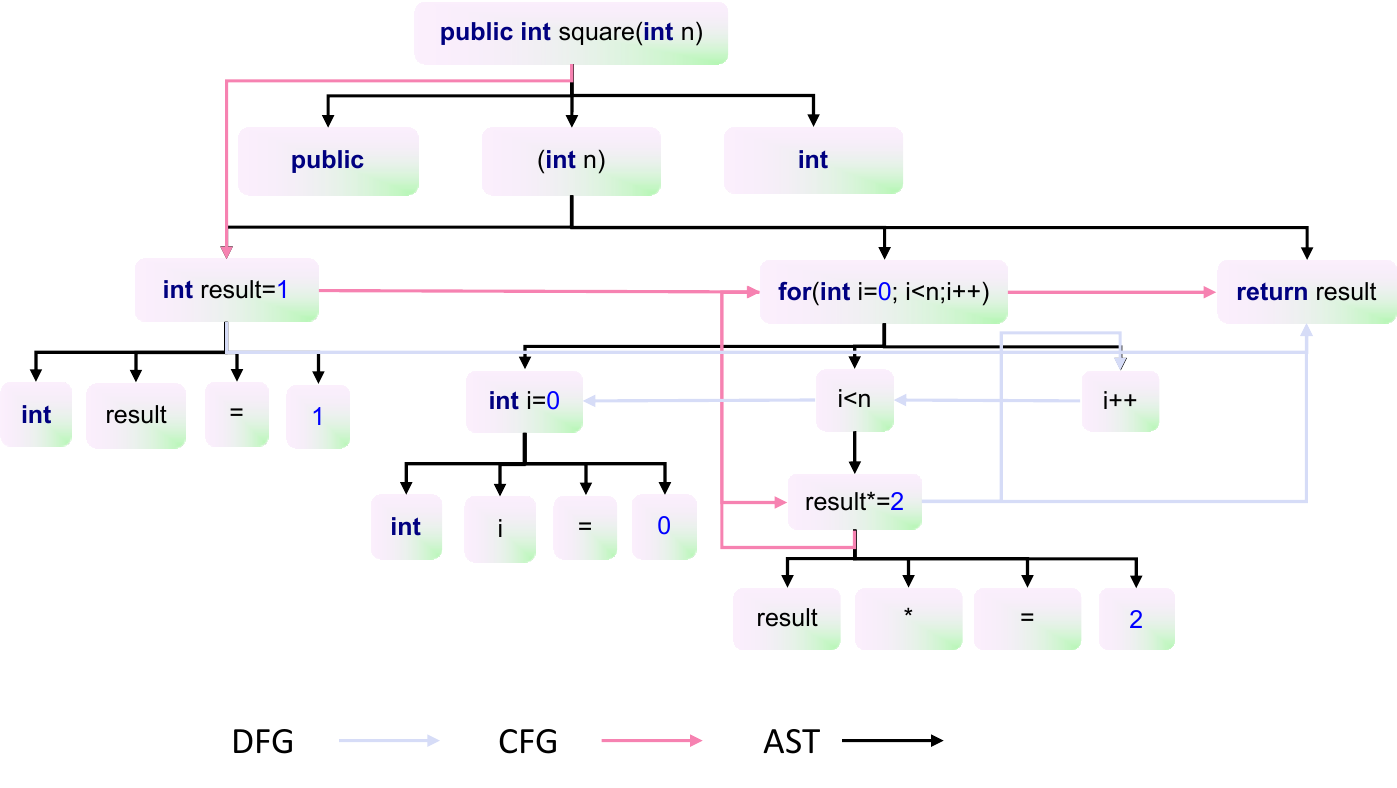}
		\caption{Structural Information}
		\label{fig:structural information}
	\end{minipage}
\end{figure*}
\subsubsection{Abstract Syntax Tree (AST)} AST is a tree representation of the abstract syntactic structure of code written in a formal language. Each node of the tree denotes a construct occurring in the code, while each edge represents the inclusion relationship between the parent node and child nodes. In the tree, every sentence links to its tokens. Specifically, a sentence forms a concatenation graph with its tokens. For example, in Fig.\ref{fig:structural information}, "int result = 1" links "int", "result", "=", and "1".
\subsubsection{Control Flow Graph (CFG)} The CFG is a graph-based representation of all pathways that a program may take during its execution. In other words, it illustrates how the program runs under various settings. All nodes belong to the branch set, which includes \texttt{switch}, \texttt{if}, \texttt{for}, and \texttt{while} statements. Each directed edge represents the program's jumps between neighboring nodes and must follow a specific jump direction. The CFG has a unique entry and exit point. The program starts at the entry point and ends at the exit point. In Fig.\ref{fig:structural information}, the \texttt{public int square(int n)} block is the entry point for the program shown in Fig.~\ref{fig:code}, and the \texttt{return result} block is the exit point. When the program reaches the \texttt{for(int i=0; i<n; i++)} block, the computer determines whether the value of the \texttt{i} variable is less than the value of the \texttt{n} variable. If the value of \texttt{i} is less than the value of \texttt{n}, the \texttt{for} block is executed; otherwise, the \texttt{return} block is executed. Thus, the \texttt{for} block is linked to both the \texttt{result*=2} block and the \texttt{return result} block.
\subsubsection{Data Flow Graph (DFG)}DFG is a graph that depicts the data dependencies between various operations, i.e., it records all variable creation and modification. In DFG, each node represents the creation or modification of variables, and each directed edge represents variables that have been modified. For example, in Fig.~\ref{fig:structural information}, the variable \texttt{result} is initialized in the \texttt{int result=1} block and is modified in the \texttt{return*=2} block. Therefore, we connect two nodes from \texttt{int result=1} to \texttt{result*=2} with a directed edge.
\subsection{MetaPath}
\label{sec:metapath}
A MetaPath is a series of object-type relations that define a new composite relation between its initial and terminating type. It is represented as a path in the form of $\theta = A_1\stackrel{R_1}{\longrightarrow}A_2\stackrel{R_2}{\longrightarrow} ... \stackrel{R_l}{\longrightarrow} A_{l+1}$, where $A_i$ is a state and $R_i$ is a composite relation between $A_i$ and $A_{i+1}$\cite{nguyen2022dsaa}. It defines a composite relation $R = R_1 \circ R_2 \circ \cdots \circ R_l$ between type $A_1$ and $A_{l+1}$, and $\circ$ denotes the composition operator on relations. The length of a MetaPath depends on the number of relations in different contexts. Predefining all potential MetaPaths of any length based on all conceivable node and edge types is challenging due to the exponential expansion of MetaPaths, increased data sparsity, and decreased training accuracy. A length-N MetaPath can be decomposed into (N-1) length-2 MetaPaths. We focus on length-2 MetaPaths\cite{10.1145/3308558.3313562,10.14778/3402707.3402736} through reflective connections between adjacent nodes to extract multiple MetaPaths, i.e., we add a reverse directed edge for a pair of nodes with a directed edge. AST, CFG, and DFG are mostly tree-like, with very few back-edges. Adding the "back" relations improves the completeness of the extracted MetaPaths and enhances the connectivity of the graph to reduce overfitting.
\subsection{Code Snippet Sequence (CSS)} CSS is a novel program language encoder that represents the semantic information of a code snippet by encoding code snippets into feature information matrices. The key to CSS is using a pre-trained program language model. Compared to traditional encoding methods, pre-trained models achieve satisfactory results without training, while significantly reducing the computing power requirements. A program language pre-training model can obtain more appropriate features than a natural language pre-training model. We calculate CSS as follows:
\begin{equation}
	C_i=model(x_i) \label{CSS}
\end{equation}
where $x_i$ is an input code snippet, $C_i$ is a representation of a code snippet and \texttt{model} represents a pre-trained program language model.
\subsection{Multi-View Self-Attention Encoder(MVSA)}
\begin{figure}[htbp]
	\includegraphics[width=50mm]{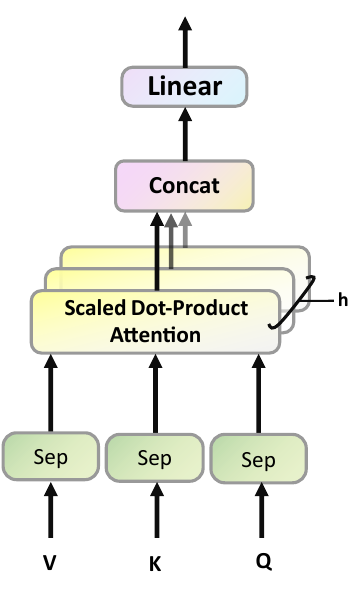}
	\centering
	\caption{Illustration of MVSA}
	\label{fig:att}
\end{figure}
After obtaining the structural information (AST, CFG, and DFG) and code semantic representation (CSS), we need to merge these representations and focus on location-specific information to extract more features. To achieve this, we design a Multi-View Self-Attention Encoder (MVSA) based on the multi-head attention mechanism. The input of MVSA consists of three matrices, $Q$, $K$, and $V$, which represent query, key, and value, respectively. Due to the self-attention mechanism, $Q$, $K$, and $V$ are the same. We compute the dot products of the query with all keys, divide each by $\sqrt{d_k}$, and apply a softmax function to obtain the weights on the values. Single-head attention is calculated as shown in Eq.~\ref{eq:sigle-att}:

\begin{equation}
	\label{eq:sigle-att}
	h_i=softmax\left(
	\frac{Q_iK_i^T}{\sqrt{d_k}}\right)
\end{equation}

where $Q_i$, $K_i$, and $V_i$ denote the $i$-th submatrix of $Q$, $K$, and $V$, respectively. $d_k$ refers to the dimension of $K$. We concatenate all $h_i$ from the scaled dot-product attention layer, calculated as shown in Eq.~\ref{eq:multi-att}:

\begin{equation}
	\label{eq:multi-att}
	G=Concat(h_i, \cdots ,h_n)W^o
\end{equation}

where $W^o$ is a weight matrix that is trained jointly with the model, and \texttt{n} is a user-defined parameter. We concatenate all $h_i$ as $H$ and multiply it with $W^o$. The resulting \texttt{G} matrix captures information from all $h_i$. Finally, we assemble four MVSAs into a quadruple self-attention layer, calculated as follows:

\begin{equation}
	\label{eq:layer}
	Q=Layer(AST, DFG, CFG, CSS)
\end{equation}

where $Q$ is a single matrix fusing of four matrices, containing both structural and semantic information.
\section{Implementation}
\label{sec:implementation}
In this section, we provide details on the implementation of JFinder, as illustrated in Fig.~\ref{fig:model}, including data preparation and module implementation.

\subsection{Generating data}
\subsubsection{Generating AST}
We employed JavaParser\footnote{http://javaparser.org/}, an open-source tool for analyzing Java code, to construct abstract syntax trees (ASTs). Using the \texttt{parse} module, we generated an AST for a given code snippet and outputted a DOT file containing edge and node information. For instance, \texttt{n0 -> n1} denotes an edge from the 0th node to the 1st node. Based on this information, we constructed the AST adjacency matrix.

\subsubsection{Generating CFG and DFG}
We extracted C++ code using tree-sitter-c~\footnote{https://tree-sitter.github.io/tree-sitter/}, a parser-generating tool and incremental parsing library that produces concrete syntax trees for source files and efficiently updates the syntax trees as the source files change. We then created a Node class to store the current node's header nodes, end node, and next node, enabling straightforward traversal of all nodes and their associated nodes. Next, we used lists to store related nodes, added directed edges between them based on conditional expressions, and created the control flow graph (CFG) adjacency matrix using these directed edges. For example, for node 4 with two child nodes 5 and 6, we set matrix entries $M(4,5)=1$ and $M(4,6)=1$.

The data flow graph (DFG) adjacency matrix generation process is similar to that of the CFG adjacency matrix. The only difference is that we added directed edges based on data flow instead of conditional expressions.

\subsubsection{Generating CSS}
We employed the HuggingFace transformer library~\cite{wolf-etal-2020-transformers}, which includes a framework of pre-trained models. We loaded the UniXcoder model~\cite{guo2022unixcoder} based on this framework, a unified cross-modal pre-trained model for programming languages that supports both code-related understanding and generation tasks. Before generating code snippet embeddings, we tokenized code snippets into token sequences. We used the UniXcoder tokenizer to process our datasets. However, the tokenizer's performance was suboptimal, as it divided \texttt{LF\_NORMAL} into \texttt{LF}, \texttt{\_}, and \texttt{NORMAL}. In our experiments, we found that incorrect code tokenization led to a significant decrease in our model's accuracy. To address this issue, we expanded the UniXcoder's vocabulary by traversing our datasets and recording words not present in the UniXcoder's vocabulary as a special word list, which we then added to the UniXcoder. For each source code snippet, the UniXcoder outputted a code semantic snippet (CSS) embedding matrix containing its semantic information. We should note that the source code length cannot exceed 512 tokens, and any source code exceeding this length must be truncated.
\subsection{module implementation}
\subsubsection{Metapath}
As shown in Fig.\ref{fig:model} and discussed in Sec.\ref{sec:metapath}, we needed to add reversed edges for each pair of nodes connected by a directed edge. To customize our metapaths, we wrote a Python program to add metapaths to the AST, CFG, and DFG adjacency matrices, as shown below:
\begin{algorithm}[]
	\caption{Metapath module}
	\LinesNumbered
	\KwIn{Matrix $x_i \in AST, CFG, DFG$}
	\KwOut{Matrix $y_i$}
	\For{i=0 i$<x_i$.1st-dimention i++}{
		\For{q=0 q$<x_i$.2nd-dimention q++}{
			\If{$x_i$(i,q)=1}{
				$x_i$(q,i)$\leftarrow$1
			}
		}
	}
\end{algorithm}
\subsubsection{Multi-View Self-Attention Encoder (MVSA)}
We implemented the MVSA using Keras~\cite{chollet2015keras} based on TensorFlow~\cite{abadi2016tensorflow}, an advanced neural network library for building and training deep learning models that offers an easy-to-use high-level interface for constructing, training, and evaluating deep learning models. We created a custom layer containing four MVSAs, named Quad Self-Attention Layer. The outputs of the Quad Self-Attention Layer were fed into a convolutional layer and a fully connected layer, which predicted whether a code snippet was vulnerable. Our model employed cross-entropy as the loss function, calculated as follows:
\begin{equation}
	\mathcal{L}_{CE}=\sum_{i=0}^{N} y_{i} \log p_{i}+\left(1-y_{i}\right) \log \left(1-p_{i}\right)\label{eq_ce},
\end{equation}
where $y_i$ represents the true label and $p_i$ is the probability of the label predicted by the model.
\section{Evaluation}
\label{sec:eval}
\subsection{Experimental Setup}
In this section, we describe the primary performance indicators for evaluating the model, the datasets, and the experimental environment.

\subsubsection{Performance Metrics}
We employ F1 scores and accuracy to assess our model. These two metrics are widely used for evaluating model performance. We provide a brief explanation of these two metrics.\par
\textbf{Accuracy: }Accuracy is the ratio of the number of samples correctly predicted by the model to the total number of samples.\par
\textbf{Precision: }Precision is the ratio of the number of files correctly classified as vulnerable to the number of files classified as vulnerable.\par
\textbf{Recall: }Recall is the ratio of the number of files correctly classified as vulnerable to the total number of genuinely vulnerable files.\par
\textbf{F1 Scores: }F1 scores are the harmonic mean of precision and recall, calculated as follows:
\begin{equation}
	2 \times \frac{Recall \times Precision}{Recall + Precision}
\end{equation}

\subsubsection{Dataset}
We use thirteen datasets to evaluate our model, including CWE datasets and the PROMISE dataset\footnote{http://openscience.us/repo/defect/}. We selected CWE datasets from the NIST Software Assurance Reference Dataset\footnote{https://samate.nist.gov}, which is the software certification and evaluation division of the National Institute of Standards and Technology (NIST). This division is responsible for researching and developing software security assessment and certification standards. We chose six CWE datasets from SARD (shown in Table ~\ref{table:cwe_dataset}), including CWE 15, CWE 23, CWE 36, CWE 89, CWE 259, and CWE 606. These datasets consist of simplified vulnerability source codes derived from real software but have been artificially modified and patched. The PROMISE Repository is a publicly available repository specializing in software engineering research datasets. We selected the following Java project datasets: Camel, jEdit, Lucene, POI, Synapse, Xalan, and Xerces, which contain real software vulnerabilities without manual modification (shown in Table ~\ref{table:promise_dataset}).

\begin{table*}[htbp]
	\centering
	\caption{CWE Dataset Information}
	\label{table:cwe_dataset}
	\begin{tabular}{ccccc|cc}
		\hline
		Project & Training Set & Validation Set & Test Set & Total & Vul  & Non-Vul \\ \hline
		CWE259  & 218          & 27             & 28       & 273   & 111  & 162     \\
		CWE606  & 775          & 97             & 97       & 969   & 333  & 636     \\
		CWE36   & 1046         & 131            & 131      & 1308  & 660  & 648     \\
		CWE15   & 1046         & 131            & 131      & 1308  & 660  & 648     \\
		CWE23   & 1046         & 131            & 131      & 1308  & 660  & 648     \\
		CWE89   & 3876         & 484            & 485      & 4845  & 1665 & 3180    \\ \hline
	\end{tabular}
\end{table*}

\begin{table*}[htpb]
	\centering
	\caption{PROMISE Dataset Information}
	\label{table:promise_dataset}
	\begin{tabular}{ccccc|cc}
		\hline
		Project & Training Set & Validation Set & Test Set & Total & Vul & Non-Vul \\ \hline
		camel   & 2156         & 269            & 270      & 2695  & 562 & 2133    \\
		jedit   & 1309         & 164            & 164      & 1637  & 277 & 1360    \\
		lucene  & 600          & 75             & 75       & 750   & 437 & 313     \\
		poi     & 1086         & 136            & 136      & 1358  & 706 & 652     \\
		synapse & 494          & 62             & 62       & 618   & 157 & 461     \\
		xalan   & 1831         & 229            & 229      & 2289  & 893 & 1396    \\
		xerces  & 835          & 104            & 105      & 1044  & 214 & 830     \\ \hline
	\end{tabular}
\end{table*}

\subsubsection{Environment Configuration}
Our experiment's hardware configuration was executed on a multi-core computing server featuring a 16-core 2.10 GHz Intel Xeon CPU and an NVIDIA 3090 GPU. The server has 256 GB of RAM and 24 GB of VRAM. The software configuration includes TensorFlow v2.7.0 and Keras v2.7.0 running on Windows 10. For MVSA, we set the head number to 4 for CSS, CFG, DFG, and AST. For the convolutional and fully connected layers, we use the Adam optimizer with a learning rate of 1e-5 and a batch size of 16. The overall training process takes approximately 5 minutes for each dataset. The final trained model has over 4,000,000 hyperparameters.

\subsection{Baseline Methods}
\subsubsection{Traditional~\cite{tradictional}}The traditional method employs a Logistic Regression classifier based on 20 features.
\subsubsection{DBN~\cite{DBN}}DBN utilizes a Deep Belief Network on source code to extract semantic features for defect prediction.
\subsubsection{DBN+\cite{DBN+DP-CNN}}DBN+, an improved version of DBN, combines semantic features with traditional features.
\subsubsection{CNN\cite{cnn}}CNN treats source codes as natural languages, with Word2Vec used for embedding initialization. It employs a CNN to extract features from source codes.
\subsubsection{Defect Prediction via Convolutional Neural Network (DP-CNN)\cite{DBN+DP-CNN}}DP-CNN uses a CNN for automated feature generation from source code while preserving semantic and structural information. It employs word embedding and combines CNN-learned features with traditional hand-crafted features to further improve defect prediction.
\subsubsection{Improved CNN\cite{improvecnn}}The improved CNN model can learn semantic representations from source-code ASTs. It enhances global pattern capture ability and improves the model for better generalization.
\subsubsection{Achilles~\cite{Achilles}}Achilles is a Java source code security vulnerability detection tool based on LSTM RNN models. It can be trained using vulnerability source code datasets, analyze Java programs, and predict security vulnerabilities at the method level.
\subsubsection{Intelligent Sentence-level Vulnerability Self-Detection Framework(ISVSF)~\cite{ISVSF}}ISVSF considers the syntax characteristics of Java and adopts sentence-level method representation and pattern exploration.

\subsection{Experimental Results}
\begin{figure*}[!htb]
	\centering
	\subfloat[Validation Accuracy\label{fig:val_acc}]{
		\includegraphics[width=3.1in]{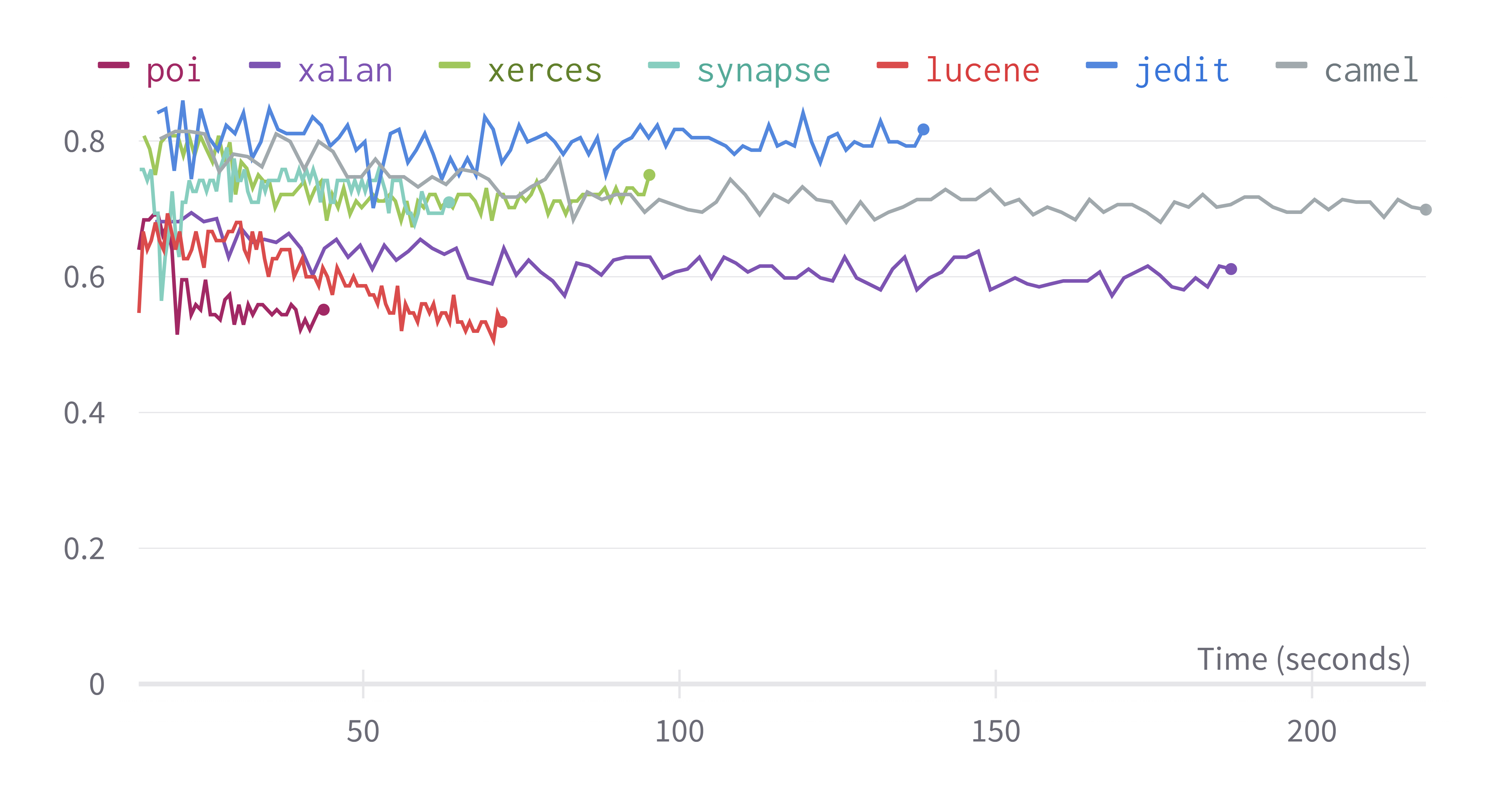}
	}
	\subfloat[Validation Precision\label{fig:val_pre}]
	{
		\includegraphics[width=3.1in]{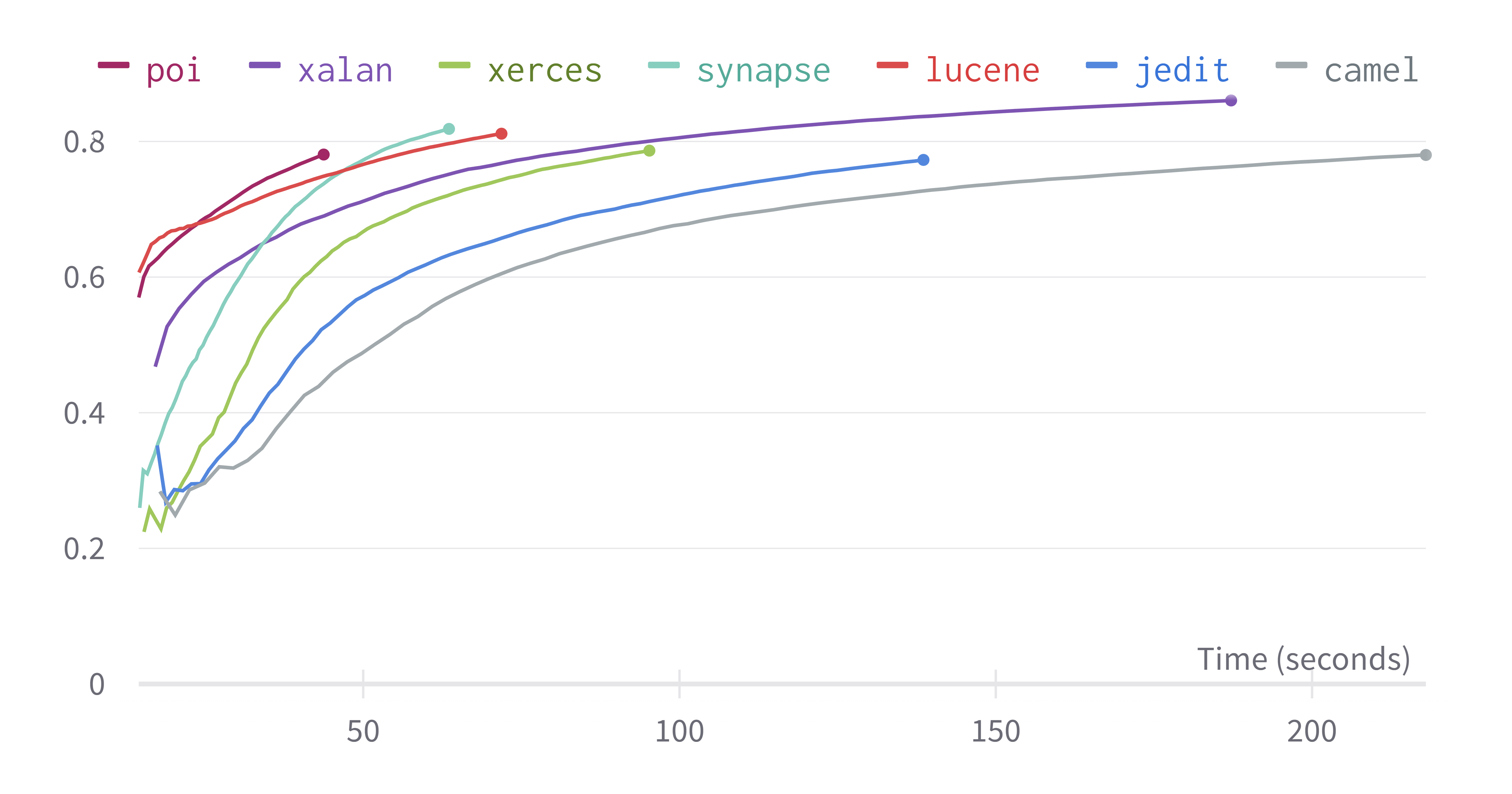}
	}

	\subfloat[Validation Recall\label{fig:val_recall}]
	{
		\includegraphics[width=3.1in]{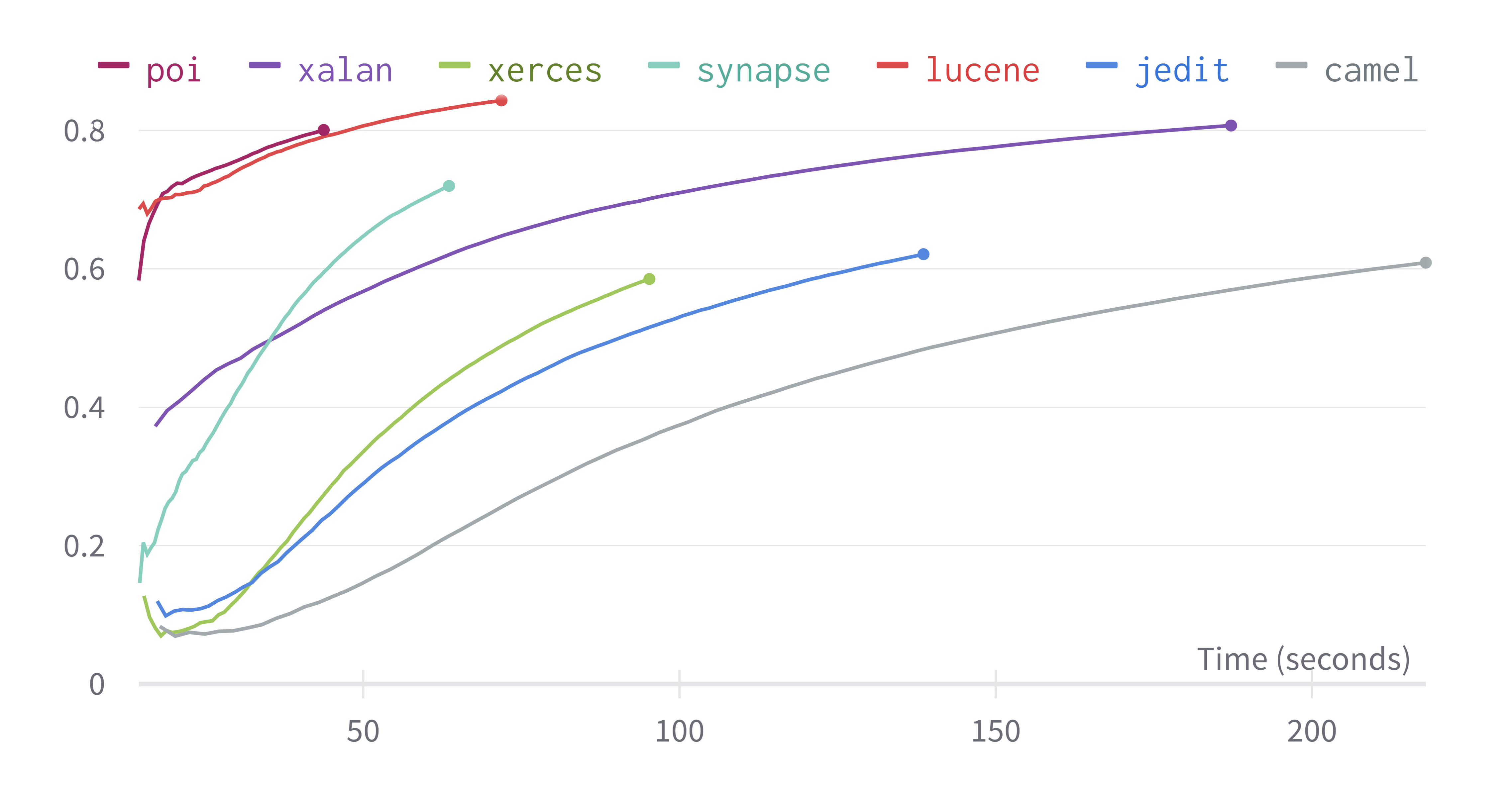}
	}
	\subfloat[Validation F1\label{fig:val_f1}]
	{
		\includegraphics[width=3.1in]{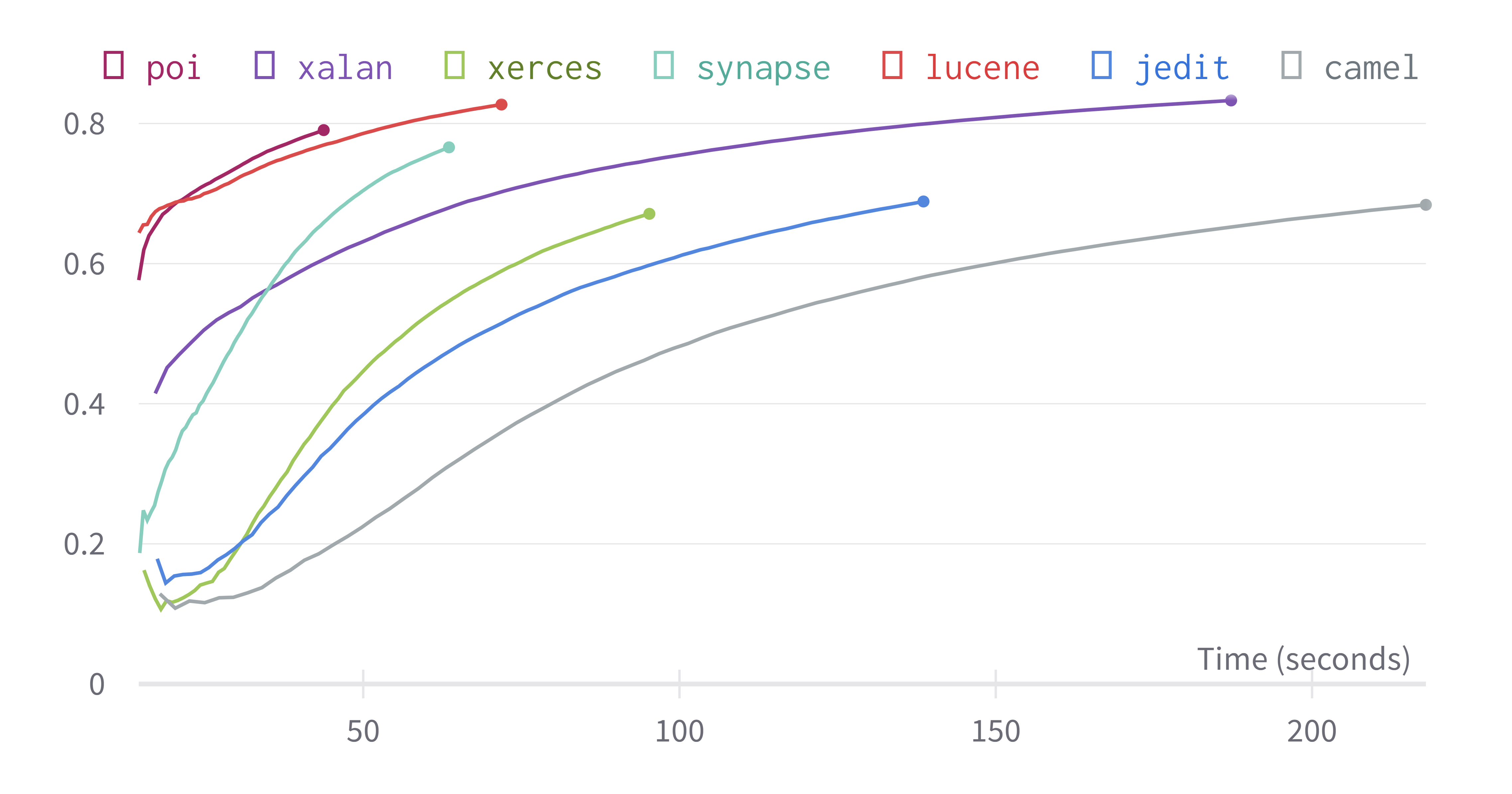}
	}
	\caption{Training Information}
	\label{fig:val_inf}
\end{figure*}

We evaluate our model on thirteen datasets shown in Table ~\ref{table:cwe_dataset} and Table ~\ref{table:promise_dataset}. JFinder outperforms all baselines in both conventional and highly complex datasets, demonstrating that our model represents the state of the art in Java vulnerability detection. According to the experimental results (shown in Table ~\ref{table:cwe_acc} and Table ~\ref{table:promise_f1}), we summarize the following findings:

\begin{itemize}
	\item \textbf{Code structural information improves vulnerability identification performance.} Compared to Achilles and ISVSF, which incorporate ASTs without other structural information, our model's accuracy is more than 5\% higher. Additionally, ISVSF outperforms Achilles in some datasets due to its inclusion of ASTs.
	\item \textbf{Pre-trained code semantic models enhance the ability of models to learn code representation semantics.} None of the baseline methods use a pre-training mechanism, resulting in their poor performance on real datasets. JFinder outperforms them by up to 35\% in F1 scores.
	\item \textbf{Quadruple self-attention layer extracts code vulnerability patterns, and aggregates structural information and semantic representation of source code.} Comparing CNN, DP-CNN, and improved CNN, JFinder's F1 scores are more than 25\% higher, indicating the positive effect of the quadruple self-attention layer.
	\item \textbf{JFinder performs better on highly complex datasets compared to conventional datasets.} Although JFinder surpasses all baseline methods in evaluations with commonly used datasets, its distinctiveness is not readily apparent. This is primarily due to the robust nature of the baseline methods, as well as researchers' adeptness in identifying elementary vulnerabilities. To underscore JFinder's superior performance, we employed a real-world dataset, which revealed JFinder to be 25\% more effective in terms of F1 scores than the baseline method, signifying industry-ready performance standards. \textcolor{blue}{Notwithstanding, we discerned notable disparities in the model's performance across different datasets. The CWE dataset, being artificially created, comprises vulnerabilities that are relatively simplistic and easy to detect. Conversely, the PROMISE dataset, derived from open-source projects, encompasses more concealed and challenging-to-detect vulnerabilities. Yet, in the face of such complexity, our model demonstrates commendable performance.}
	\item \textbf{JFinder achieves satisfactory performance with a short training time.} As seen in Fig.~\ref{fig:val_inf}, accuracy and F1 scores increase dramatically in a short period of time. The model is trained in less than five minutes. 
\end{itemize}

\begin{table*}[]
	\centering
	\caption{Experimental Results on PROMISE}
	\label{table:promise_f1}
	\begin{tabular}{llllllll}
		\hline
		\multirow{2}{*}{\textbf{Method}} & \textbf{camel} & \textbf{jEdit} & \textbf{synpase} & \textbf{xerces} & \textbf{lucene} & \textbf{xalan} & \textbf{poi} \\ \cline{2-8} 
		& F1             & F1             & F1               & F1              & F1              & F1             & F1           \\ \hline
		Traditional~\cite{tradictional}                  & 0.329          & 0.573          & 0.500            & 0.273           & 0.618           & 0.627          & 0.748        \\
		DBN~\cite{DBN}                    & 0.335          & 0.480          & 0.503            & 0.261           & 0.758           & 0.681          & 0.780        \\
		DBN+~\cite{DBN+DP-CNN}                        & 0.375          & 0.549          & 0.486            & 0.276           & 0.761           & 0.681          & 0.782        \\
		CNN ~\cite{cnn}                & 0.505          & 0.631          & 0.512            & 0.311           & 0.761           & 0.676          & 0.778        \\
		DP-CNN~\cite{DBN+DP-CNN}               & 0.508          & 0.580          & 0.556            & 0.374           & 0.761           & 0.696          & 0.784        \\
		Improved CNN~\cite{improvecnn}                  & 0.487          & 0.590          & 0.655            & 0.667           & 0.701           & 0.780          & 0.444        \\
		JFinder                          & 0.7637         & 0.8084         & 0.8233           & 0.7502          & 0.8391          & 0.8324         & 0.7899       \\ \hline
	\end{tabular}
\end{table*}

\begin{figure*}[htbp]
	\includegraphics[width=180mm]{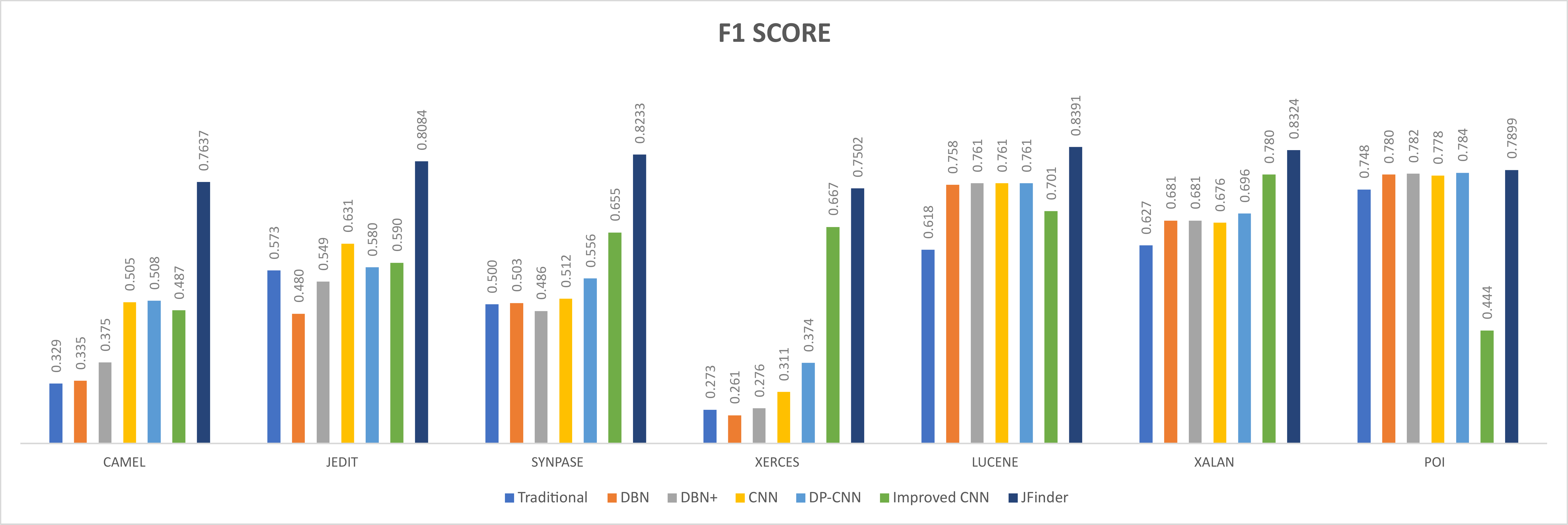}
	\centering
	\caption{F1 Score On Promise}
	\label{fig:f1}
\end{figure*}

\begin{figure*}[htbp]
	\includegraphics[width=120mm]{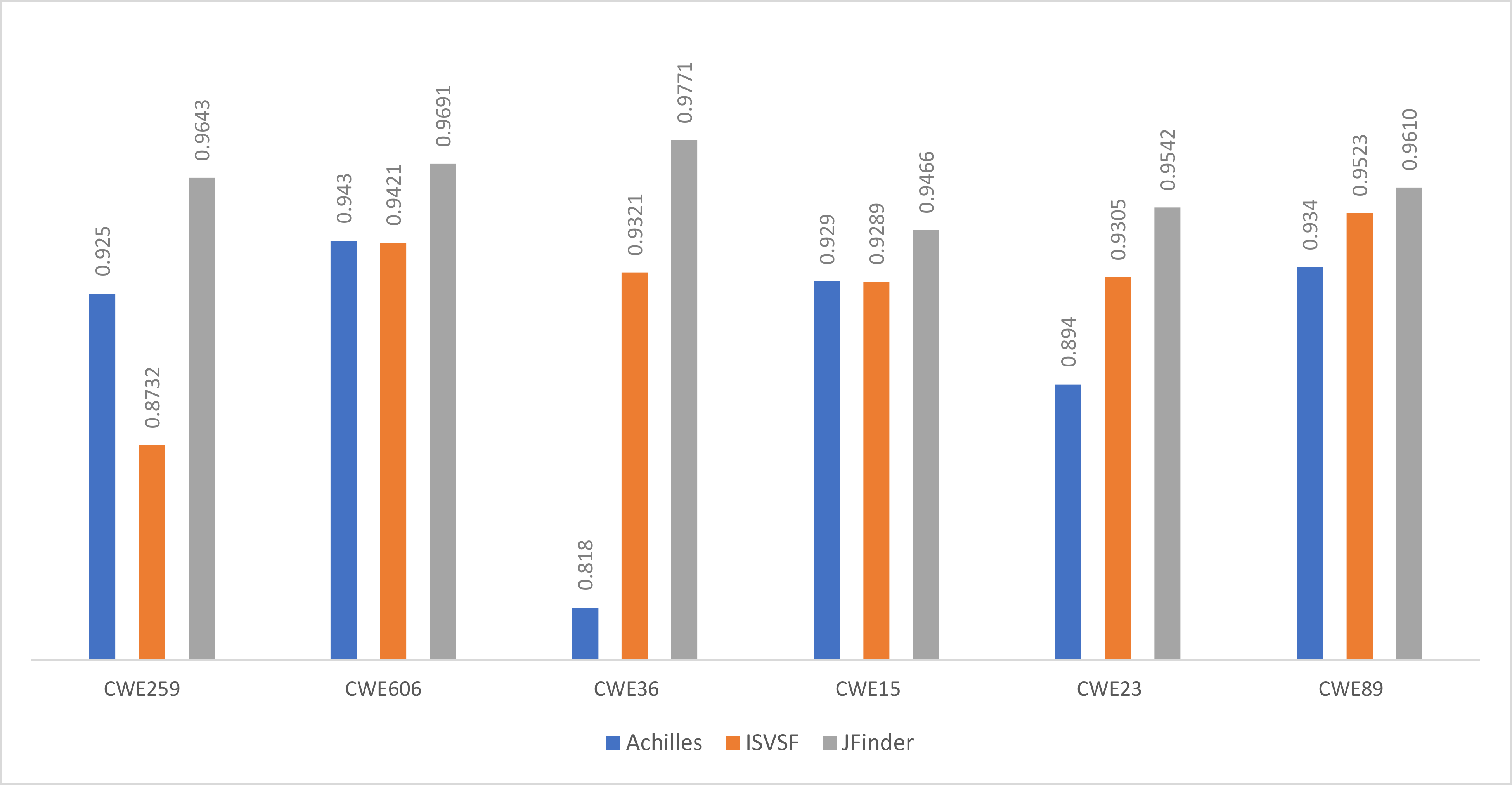}
	\centering
	\caption{Accuracy on CWE}
	\label{fig:acc}
\end{figure*}

\begin{table*}[htbp]
	\centering
	\caption{Experimental Results on CWE}
	\label{table:cwe_acc}
	\begin{tabular}{ccccccc}
		\hline
		\multirow{2}{*}{\textbf{Method}} & \textbf{CWE259} & \textbf{CWE606} & \textbf{CWE36} & \textbf{CWE15} & \textbf{CWE23} & \textbf{CWE89} \\ \cline{2-7} 
		& ACC             & ACC             & ACC            & ACC            & ACC            & ACC            \\ \hline
		Achilles~\cite{Achilles}                      & 0.925           & 0.943           & 0.818          & 0.929          & 0.894          & 0.934          \\
		ISVSF ~\cite{ISVSF}                       & 0.8732          & 0.9421          & 0.9321         & 0.9289         & 0.9305         & 0.9523         \\
		JFinder                          & 0.9643          & 0.9691          & 0.9771         & 0.9466         & 0.9542         & 0.9610         \\ \hline
	\end{tabular}
\end{table*}

\subsection{Ablation Study}
We conducted an ablation study to explore the impact of various components of our model on its performance. We utilized the PROMISE datasets to test the performance of the model components.

\textbf{Structural Information}

In the ablation experiment, we removed one of the structural information matrices, as shown in Table \ref{table:ablation}. According to the table, removing any of the structural information matrices resulted in a decrease in F1 scores, indicating that structural information provides a significant number of features to the model. Upon analyzing the results further, we found that CFG and DFG are more critical than AST. When either CFG or DFG was removed, the F1 scores dropped more substantially. However, for the \texttt{poi} dataset, they played similar roles. Thus, we infer that CFG and DFG provide the model with data flow and control flow information containing more vulnerability patterns. In summary, it is unwise to remove any of the graphs, even if they appear less important.

\textbf{Semantic Information and Pre-trained Model}

Based on Table \ref{table:ablation}, we conclude that the semantic information of the source code is highly important, as it is the origin of most vulnerability features. Removing CSS resulted in a significant drop in F1 scores to 0.51. Evidently, CSS is the most critical information. Our novel quadruple self-attention layer design and pre-training mechanism play the most significant role. If this component is removed, the model becomes inoperative.
\begin{table*}[htpb]
	\centering
	\caption{Ablation Study}
	\label{table:ablation}
	\begin{tabular}{cccccccc}
		\hline
		\multirow{2}{*}{\textbf{Method}} & \textbf{camel} & \textbf{jEdit} & \textbf{synpase} & \textbf{xerces} & \textbf{lucene} & \textbf{xalan} & \textbf{poi} \\ \cline{2-8} 
		& F1             & F1             & F1               & F1              & F1              & F1             & F1           \\ \hline
		JFinder                          & 0.7637         & 0.8084         & 0.8233           & 0.7502          & 0.8391          & 0.8324         & 0.7899       \\
		Jfinder w/o AST                  & 0.7204         & 0.7904         & 0.8164           & 0.7443          & 0.8273          & 0.8085         & 0.7730       \\
		JFinder w/o CFG                  & 0.7343         & 0.7551         & 0.8199           & 0.6982          & 0.6292          & 0.7028         & 0.7603       \\
		JFinder w/o DFG                  & 0.5963         & 0.7378         & 0.7763           & 0.7342          & 0.5814          & 0.7283         & 0.7764       \\
		JFinder w/o CSS                  & 0.4895         & 0.5933         & 0.7111           & 0.5159          & 0.6882          & 0.6760         & 0.7166       \\ \hline
	\end{tabular}
\end{table*}

\subsection{Case Study}
To further assess the robustness and intelligence of JFinder, we examined four representative vulnerability cases in the CWE dataset. Our model correctly identified each case. First, we analyzed the vulnerability specifications. Next, we manually patched the vulnerabilities and input them into our model to determine if it no longer flagged them, checking its ability to deeply understand the meaning of the vulnerabilities.
\subsubsection{Case 1}
\begin{figure}[!htb]
	\centering
	\subfloat[vulnerability\label{fig:case1 vul}]{

		\includegraphics[width=3.1in]{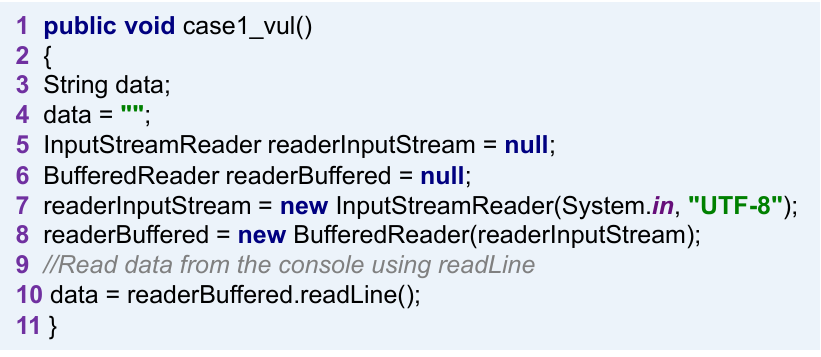}
				
	}

	\subfloat[fix\label{fig:case1 fix}]
	{

		\includegraphics[width=1.8in]{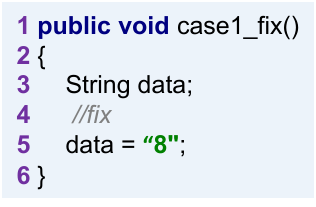}
				
	}

	\caption{Case 1}
	\label{fig:case1}
\end{figure}

In Case 1 (Fig.\ref{fig:case1}), the program did not check input and read data from the console using \texttt{readLine()}, potentially causing a denial of service or other consequences due to excessive looping. The data originated from a malicious source, as shown in Fig.\ref{fig:case1 vul} Line 10. To remediate the vulnerability (Fig.~\ref{fig:case1 fix}), we replaced user-controlled data for loop conditions with a hardcoded string. Afterward, JFinder no longer flagged this source code, indicating its ability to learn the reason behind unchecked input for loop condition problems.
	
\subsubsection{Case 2}
\begin{figure}[!htb]
	\centering
	\subfloat[vulnerability\label{fig:case2_vul}]{
		\includegraphics[width=3.1in]{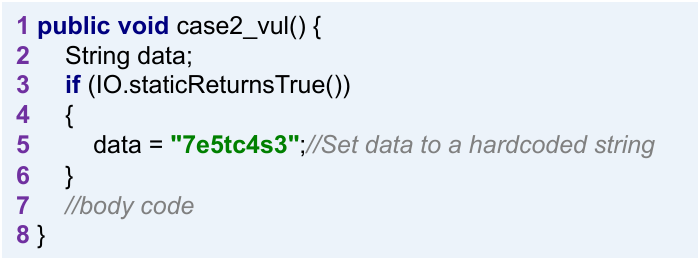}
	}

	\subfloat[fix\label{fig:case2_fix}]
	{
		\includegraphics[width=3.1in]{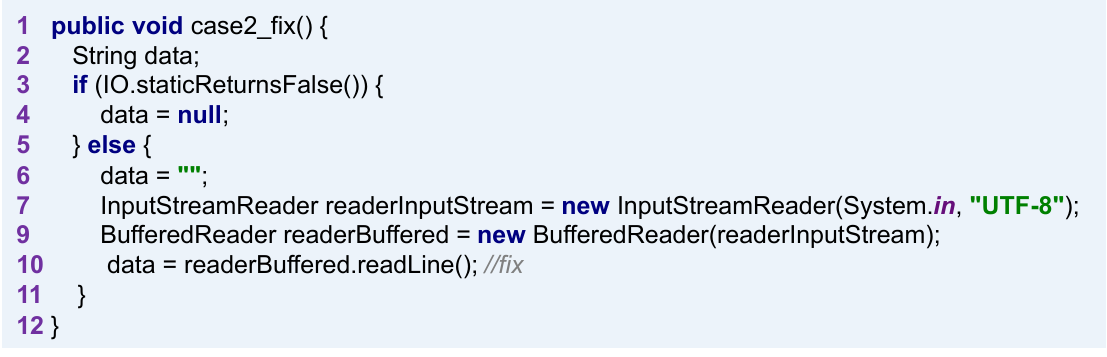}
	}
	\caption{Case 2}
	\label{fig:case2}
\end{figure}

In Case 2 (Fig.\ref{fig:case2}), the code snippet exhibited a vulnerability involving the use of a hard-coded password. A hard-coded password can lead to significant authentication failures that may be difficult for system administrators to detect and fix. In extreme cases, administrators may be forced to disable the product entirely. As seen in Fig.\ref{fig:case2_vul} Line 5, the program established a hard-coded password, suggesting that if such passwords are used, malicious users are likely to gain access through the account in question. To patch the vulnerability, we set \texttt{data} via external input, as shown in Fig.~\ref{fig:case2_fix}. After testing, JFinder no longer flagged the program as vulnerable.

\subsubsection{Case 3}
\begin{figure}[!htb]
	\centering
	\subfloat[vulnerability\label{fig:case3_vul}]{
		\includegraphics[width=3.1in]{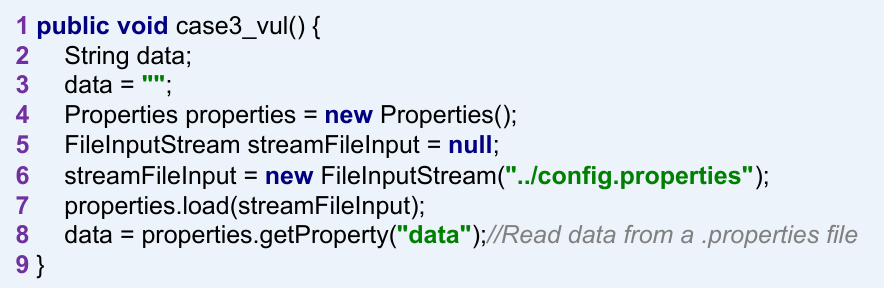}
	}

	\subfloat[fix\label{fig:case3_fix}]
	{
		\includegraphics[width=3.1in]{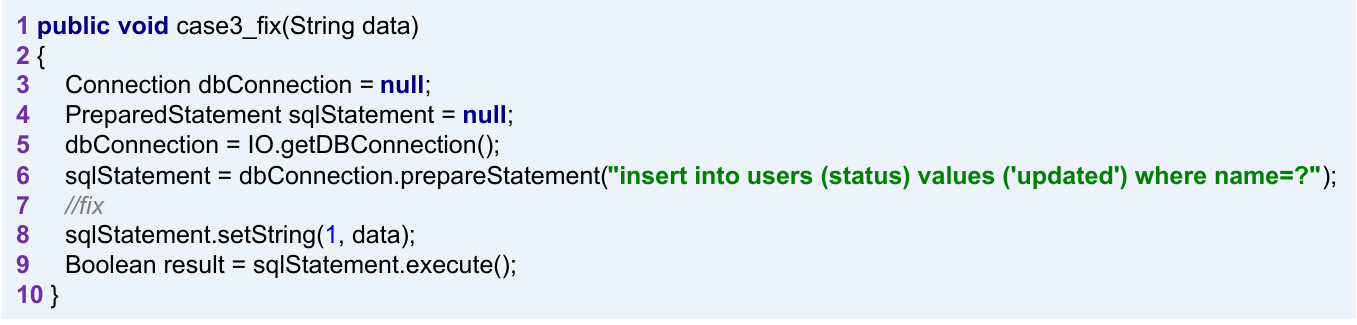}
	}
	\caption{Case 3}
	\label{fig:case3}
\end{figure}

In Case 3 (Fig.\ref{fig:case3}), a vulnerability appeared in Line 7, as depicted in Fig.\ref{fig:case3_vul}. In Line 8, the program read data from a properties file, leading to SQL injection. Without sufficient removal or quoting of SQL syntax in user-controllable inputs, the generated SQL query can cause those inputs to be interpreted as SQL rather than ordinary user data. This can be exploited to alter query logic, bypass security checks, or insert additional statements that modify the back-end database, potentially including the execution of system commands. To fix the vulnerability, we set \texttt{data} to be passed through the function instead of being read from the properties file. Afterward, JFinder considered the code snippet non-vulnerable.

\subsubsection{Case 4}
\begin{figure}[!htb]
	\centering
	\subfloat[vulnerability\label{fig:case4_vul}]{
		\includegraphics[width=3.1in]{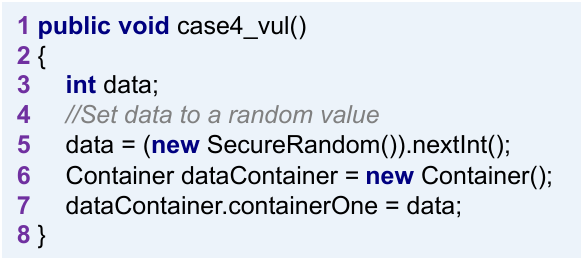}
	}

	\subfloat[fix\label{fig:case4_fix}]
	{
		\includegraphics[width=3.1in]{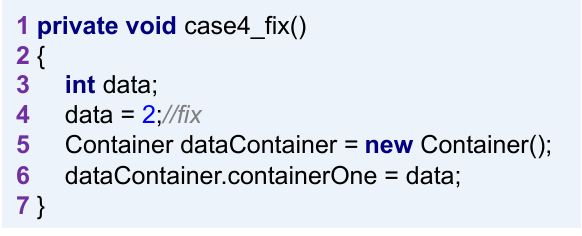}
	}
	\caption{Case 4}
	\label{fig:case4}
\end{figure}

In Case 4 (Fig.\ref{fig:case4}), a vulnerability arose from allocating memory based on an untrusted, large size value. The program did not ensure that the size was within expected limits, allowing arbitrary amounts of memory to be allocated. As shown in Fig.\ref{fig:case4_vul} Line 5, the program set \texttt{data} to a random value. To address this issue, we used a hardcoded number that would not cause underflow, overflow, divide by zero, or loss-of-precision problems. After fixing it in Fig.~\ref{fig:case4_fix}, JFinder no longer deemed it a vulnerability.

\section{Related Works}
\label{sec:related}

The automation of software vulnerability identification is a topic of great interest for researchers, who continue to develop new techniques to detect vulnerabilities. Approaches range from traditional methods that manually establish vulnerability rules, to machine learning techniques that determine vulnerabilities based on features, and to deep learning that learns vulnerability patterns.

\subsection{Traditional Detection Methods}
Traditional detection methods rely on known vulnerability rules to detect vulnerabilities, using manual code reviews and automated code scanners. Kaur \emph{et al.} introduced five static code analysis tools for vulnerability detection in C/C++ and Java~\cite{KAUR20202023}. Flawfinder~\cite{Flawf8239638:online} is designed to detect vulnerabilities in C/C++ source code, generating a list of vulnerabilities for the program sorted by risk level. RATS~\cite{Googl8714453:online} is a security vulnerability auditing tool for C/C++ source code that can detect issues such as buffer overflows. SPOTBUGS~\cite{SpotB2548343:online} is a program that uses static analysis to identify bugs in Java code, checking for more than 400 bug patterns. PMD~\cite{PMD6397148:online} is an open-source source analysis tool that employs rule sets to find common coding errors, irregular code, and potential vulnerabilities. \textcolor{blue}{Peguero \emph{et al.} analyzed Electron application security, revealing potential front-to-back-end attack escalation. They proposed framework modifications and an IDE plugin for early vulnerability fixes. Their studies confirmed the effectiveness of the plugin, as applications ceased to be exploitable post-fix.\cite{PEGUERO2021100032}}

\subsection{Machine Learning-based Methods}
When integrating various types of conventional hand-crafted features, the challenge remains how to effectively combine these features. Machine learning-based methods can address this issue by performing simple classification on manually extracted features with better performance than traditional methods. Yang \emph{et al.} proposed a deep learning model for just-in-time defect prediction~\cite{7272910}, building a set of expressive features from a set of initial change features using a deep belief network algorithm and constructing a classifier based on the selected features. Chen \emph{et al.} proposed a model~\cite{2018chen} capable of identifying SQL injection vulnerabilities by processing HTTP request text data using word2vec and classifying processed samples with the SVM algorithm. Al-Yaseen \emph{et al.} suggested a multi-level hybrid intrusion detection model that employs support vector machines and extreme learning machines to enhance the efficiency of detecting known and unknown attacks~\cite{10.1016/j.eswa.2016.09.041}. Ren \emph{et al.} introduced DVCMA~\cite{DVCMA}, a software vulnerability detection method based on clustering and model analysis that applies clustering techniques to mine patterns from vulnerability sequences. \textcolor{blue}{Peguero \emph{et al.} analyzes cross-site request forgery vulnerabilities in several server-side JavaScript frameworks. Utilizing automated static analysis, the security efficacy of each is evaluated. Based on these insights, recommendations for more secure application development are provided.\cite{PEGUERO2021100035}}

\subsection{Deep Learning-based Methods}
Manual inspection of source code or manual extraction of features from the source code is time-consuming. Deep learning-based methods can automatically extract vulnerability patterns and classify them based on the input source code's features. Zhan \emph{et al.} proposed ISVSF~\cite{ISVSF}, an intelligent sentence-level vulnerability self-detection framework that considers Java syntax characteristics and adopts sentence-level method representation and pattern exploration. Malhotra \emph{et al.} suggested an improved CNN~\cite{improvecnn}, a modified CNN algorithm that combines CNN-based layers into one and then applies a concatenate algorithm under SVM. Saccente \emph{et al.} introduced Project Achilles~\cite{Achilles}, a Java source code security vulnerability detection tool built upon LSTM RNN models. Lin \emph{et al.} proposed VulEye~\cite{app13020825}, a graph neural network vulnerability detection approach for PHP applications that utilizes program dependence graphs as input and is trained with a graph neural network model containing three stack units. \textcolor{blue}{Zheng \emph{et al.} presented CodeGeeX, a multilingual, 13 billion-parameter model surpassing peers in code generation and translation on HumanEval-X. The model enhances coding efficiency for 83.4\% of users through developed extensions. In September 2022, all associated CodeGeeX resources were publicly released. \cite{zheng2023codegeex} Raymond Li \emph{et al.} presented StarCoder and StarCoderBase, Code LLMs with 15.5B parameters and 8K context length. Trained on The Stack's 1 trillion tokens, StarCoder, a fine-tuned StarCoderBase, outperforms multilingual Code LLMs and Python-specialized models.\cite{li2023starcoder} }
\section{Conclusions and Future Research}
\label{sec:conclusion}
As modern software grows in size and complexity, ensuring stability has become a critical concern. In this paper, we introduced JFinder, a novel architecture for Java vulnerability identification based on quad self-attention and a pre-training mechanism. JFinder innovatively combines structural and semantic information through a proposed quad self-attention layer. Experimental results demonstrate that JFinder outperforms all baseline models, achieving a level of performance suitable for industrial use. JFinder's F1 scores are 25\% higher than those of the baselines, and its ACC surpasses them by 5\%. Furthermore, we conducted a case study to investigate whether the model truly understands the root causes of vulnerabilities. 
\textcolor{blue}{Looking forward, we see significant potential in exploring the use of larger language models for pre-training. The pre-training mechanism, an unsupervised learning process, enables the model to learn a wide range of syntactic and semantic patterns before fine-tuning on a specific task. The application of larger language models in this process could enhance our model's understanding of complex code structures and semantics, possibly leading to improved precision and recall in identifying vulnerabilities in Java code.}

\section*{Acknowledgment}
This study was supported by the National Key R\&D Program of China with grant No.2019YFB2102600, National Natural Science Foundation of China (62002067), the Guangzhou Youth Talent of Science (QT20220101174) and the Project of Philosophy and Social Science Planning of GuangDong GD21YGL16.

\bibliographystyle{elsarticle-num} 
\bibliography{IEEE}

\end{document}